\documentclass[12pt]{iopart}

\usepackage{iopams}  
\usepackage{graphicx}
\usepackage[breaklinks=true,colorlinks=true,linkcolor=blue,urlcolor=blue,citecolor=blue]{hyperref}

\begin{document}

\title{Superconducting Qubit-Resonator-Atom Hybrid System}

\author{Deshui Yu$^{1}$, Leong Chuan Kwek$^{1,2,3,4}$, Luigi Amico$^{1,5,6}$, \& Rainer Dumke$^{1,7}$}

\address{$^{1}$Centre for Quantum Technologies, National University of Singapore,
3 Science Drive 2, Singapore 117543, Singapore}

\address{$^{2}$Institute of Advanced Studies, Nanyang Technological University,
60 Nanyang View, Singapore 639673, Singapore}

\address{$^{3}$National Institute of Education, Nanyang Technological University,
1 Nanyang Walk, Singapore 637616, Singapore}

\address{$^{4}$MajuLab, CNRS-UNS-NUS-NTU International Joint Research Unit, UMI 3654, Singapore}

\address{$^{5}$CNR-MATIS-IMM \& Dipartimento di Fisica e Astronomia, Universit\'a Catania, Via S. Soa 64, 95127 Catania, Italy}

\address{$^{6}$INFN Laboratori Nazionali del Sud, Via Santa Sofia 62, I-95123 Catania, Italy}

\address{$^{7}$Division of Physics and Applied Physics, Nanyang Technological University, 21 Nanyang Link, Singapore 637371, Singapore}

\ead{rdumke@ntu.edu.sg}

\begin{abstract}
We propose a hybrid quantum system, where an $LC$ resonator inductively interacts with a flux qubit and is capacitively coupled to a Rydberg atom. Varying the external magnetic flux bias controls the flux-qubit flipping and the flux qubit-resonator interface. The atomic spectrum is tuned via an electrostatic field, manipulating the qubit-state transition of atom and the atom-resonator coupling. Different types of entanglement of superconducting, photonic, and atomic qubits can be prepared via simply tuning the flux bias and electrostatic field, leading to the implementation of three-qubit Toffoli logic gate.
\end{abstract}

\section{Introduction}

The most promising candidates for building a quantum computer include superconducting circuits, flying photonic qubits, and ultracold atoms. The superconducting devices~\cite{PRL:Martinis1985,Nature:Nakamura1999,Science:Mooij1999,Science:Vion2002,PRL:Martinis2002,PRA:Koch2007}, based on Josephson junctions (JJs), own the advantages of flexibility, tunability, scalability, and rapid processing quantum information, but their decoherence times are limited owing to the strong coupling to external fields~\cite{PRB:Tian2002,PRB:Martinis2003,PRB:Ithier2005,PRB:Schreier2008}. The flying qubits~\cite{Nature:Knill2001,NatPhys:Maunz2007,NatPhys:Hucul2015,NatComm:Chapman2016}, due to the nature of transmitting at the speed of light, are usually applied to implement the remote entanglement of spatially separated quantum systems. In contrast, the atoms~\cite{PRL:Monroe1995,PRL:Brennen1999,Nature:Barrett2004,Nature:Blatt2008} are characterized by the properties of indistinguishability, long-time coherence, and precise engineering quantum state. However, the logic operations based on atoms are much slower than that of superconducting qubits because of the relatively weak atom-field interaction.

A hybrid quantum structure consisting of different physical realizations would inherit the advantages of each component~\cite{PhysScr:Wallquist2009,RMP:Xiang2013,SciRep:Yu2016}. So far, the most successful hybrid system is the combination of a resonator and superconducting qubits, i.e., the so-called circuit QED~\cite{PRA:Blais2004,Nature:Wallraff2004,Nature:You2011}. Owing to the large dipole moment of a superconducting qubit and the strong field in the resonator, the intercomponent coupling well exceeds any decoherence rates, reaching the strong-coupling regime. In comparison, the strong superconducting qubit-atom and atom-resonator interactions are still in challenge due to the weak dipole moment of atom. Replacing one atom by an ensemble of atoms~\cite{PRL:Petrosyan2008,PRL:Verdu2009,PRA:Patton2013} or polar molecules with large dipole moments~\cite{PRL:Rabl2006,NaturePhysics: Andre2006} or choosing the Rydberg-Rydberg transition~\cite{PRA:Petrosyan2009,PRA:Yu2016-1}, whose energy spacing matches the typical frequency of superconducting qubits or resonator, potentially achieve the strong coupling between different components.

Combining superconducting circuits, resonator, and atoms together is of importance for transferring the quantum information between the atomic memory and the superconducting processor, where the intraresonator field serves as the intermediary bridge or data bus~\cite{Nature:Sillanpaa2007,Nature:Hofheinz2009}. This hybrid structure also provides a platform to study the entanglement and logic operations in the three-qubit space consisting of a superconducting qubit, an oscillator, and an atomic qubit, leading to the potentials of tailoring two-qubit entangled state by controlling the third qubit and linking the three-qubit unit to other distant systems potentially via flying qubits.

In this paper, based on the model established in~\cite{PRA:Yu2016-2}, we propose a hybrid system, where an $LC$ resonator inductively interacts with a flux qubit and is capacitively coupled to an atom. The qubit energy spectra are tuned by external magnetic flux bias and electrostatic field, separately controlling the flux qubit-resonator and atom-resonator interactions. Preparations of different three-qubit fully entangled states are able to be accomplished via manipulating intercomponent couplings. Toffoli logic gates can be also performed on this hybrid system.

\section{Results}

\subsection{Physical model}

\begin{figure}
\centering 
\includegraphics[width=10.0cm]{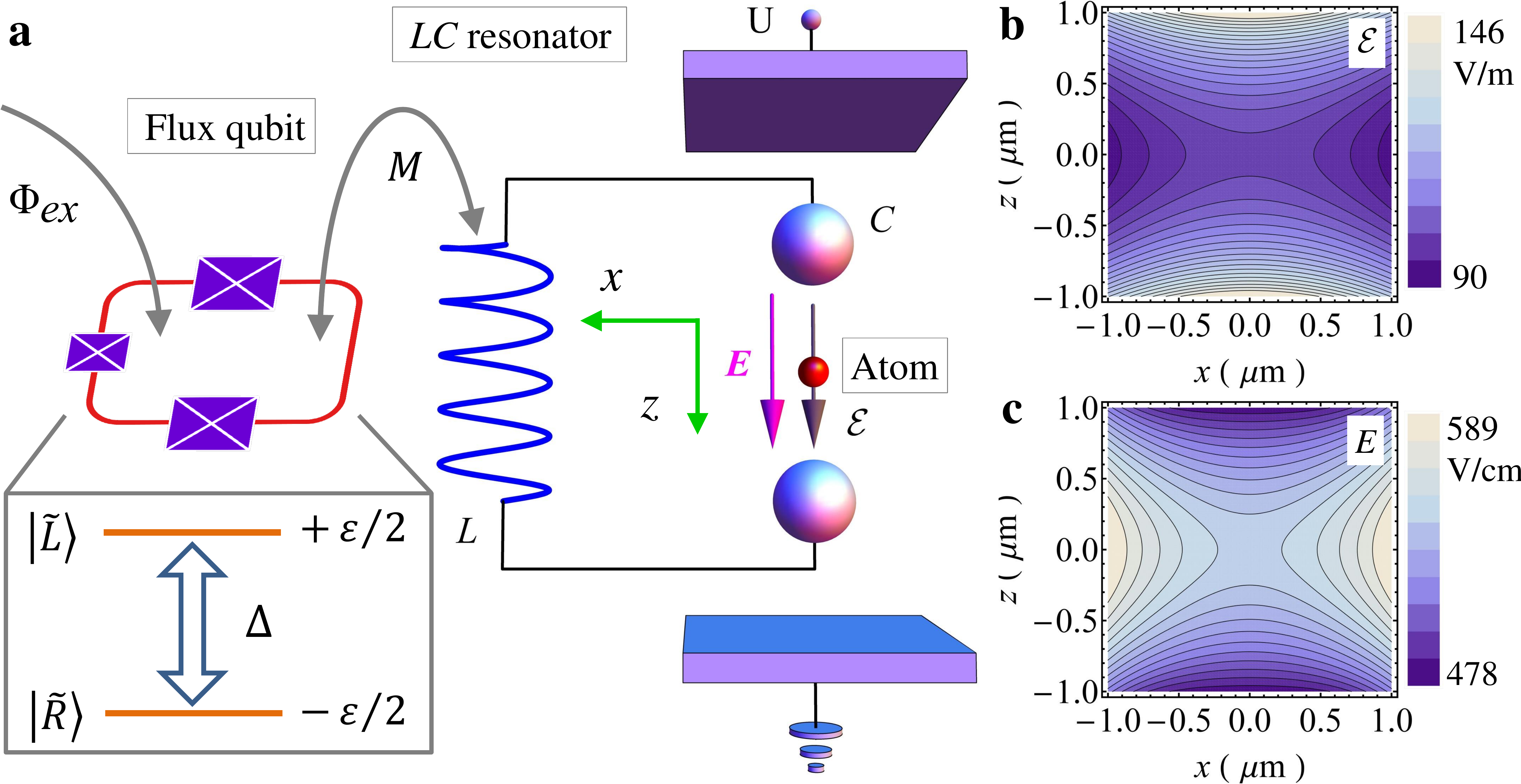}\\
\caption{(a) Hybrid system composed of a three-JJ flux qubit, a $LC$ resonator, and an atom. The flux qubit, which is biased by an external flux $\Phi_{ex}$, is formed by two circulating persistent-current states $|\tilde{L}\rangle$ and $|\tilde{R}\rangle$ with a frequency separation of $\varepsilon$ and a tunnel rate $\Delta$. The flux qubit is inductively coupled to the resonator inductor with a mutual inductance $M$. The resonator capacitor consists of two equal sized conducting spheres (radius of 3 $\mu$m and intercenter distance of 9 $\mu$m). The atom is placed in the middle point between two spheres and interacts with the oscillating intraresonator electric field with an amplitude ${\cal{E}}$ in the $z$-axis. An extra electrostatic field $E$ along the $z$ direction, generated by a parallel-plate capacitor, tunes the atomic spectrum. The distribution of ${\cal{E}}$ and $E$ in the $x-z$ plane around the atom (the origin point) are shown in (b) and (c), respectively, and the inhomogeneities are both less than 0.1\%.}\label{Fig1}
\end{figure}

We consider a superconducting $LC$ resonator operating at mK temperatures (see figure \ref{Fig1} (a)). To match the typical superconducting qubit frequency, the characteristic frequency $\omega_{0}=1/\sqrt{LC}$ is chosen to be $2\pi\times20$ GHz. The capacitor $C$, as an example, is composed of a pair of identical conducting spheres with a radius of 3 $\mu$m and an intercenter distance of 9 $\mu$m along the $z$ axis, resulting in $C=256$ aF and $L=247$ nH. Introducing the magnetic flux and capacitor charge operators $\phi=\sqrt{\hbar/(2C\omega_{0})}(b^{\dag}+b)$ and $q=i\sqrt{C\hbar\omega_{0}/2}(b^{\dag}-b)$, the resonator Hamiltonian reads
\begin{equation}
H_{LC}=\phi^{2}/(2L)+q^{2}/(2C)=\hbar\omega_{0}(b^{\dag}b+1/2),
\end{equation}
where $b^{\dag}$ and $b$ are the creation and annihilation operators with the commutation relation $[b,b^{\dag}]=1$.

A Rb Rydberg atom is placed at the midpoint between two capacitor spheres and couples to the local $z$-direction electric field produced by the charge $q$, resulting in the atom-resonator interaction operator
\begin{equation}
V_{a}=-iD{\cal{E}}(b^{\dag}-b).
\end{equation}
$D$ is the atomic dipole moment and ${\cal{E}}$ is the amplitude of the oscillating electric field inside the resonator. For the physical specification in this work, the inhomogeneity of ${\cal{E}}$ within the atomic wave pocket, i.e., $|\frac{1}{{\cal{E}}}\frac{\partial{\cal{E}}}{\partial\alpha}\delta r|$ ($\alpha=x,y,z$ and $\delta r$ is the radius of Rydberg state), is less than $10^{-3}$ (see figure \ref{Fig1} (b)) and hardly affects the atom-resonator coupling. In addition, an external electrostatic $E$ field in the $z$-axis, generated by an extra parallel-plate capacitor (see figure \ref{Fig1} (a)), is applied to tune the energy spectrum of atom. The inhomogeneity of $E$ caused by the screening effect of two capacitor spheres is also neglectable at the position of atom (see figure \ref{Fig1} (c)).

\begin{figure}
\centering
\includegraphics[width=10.0cm]{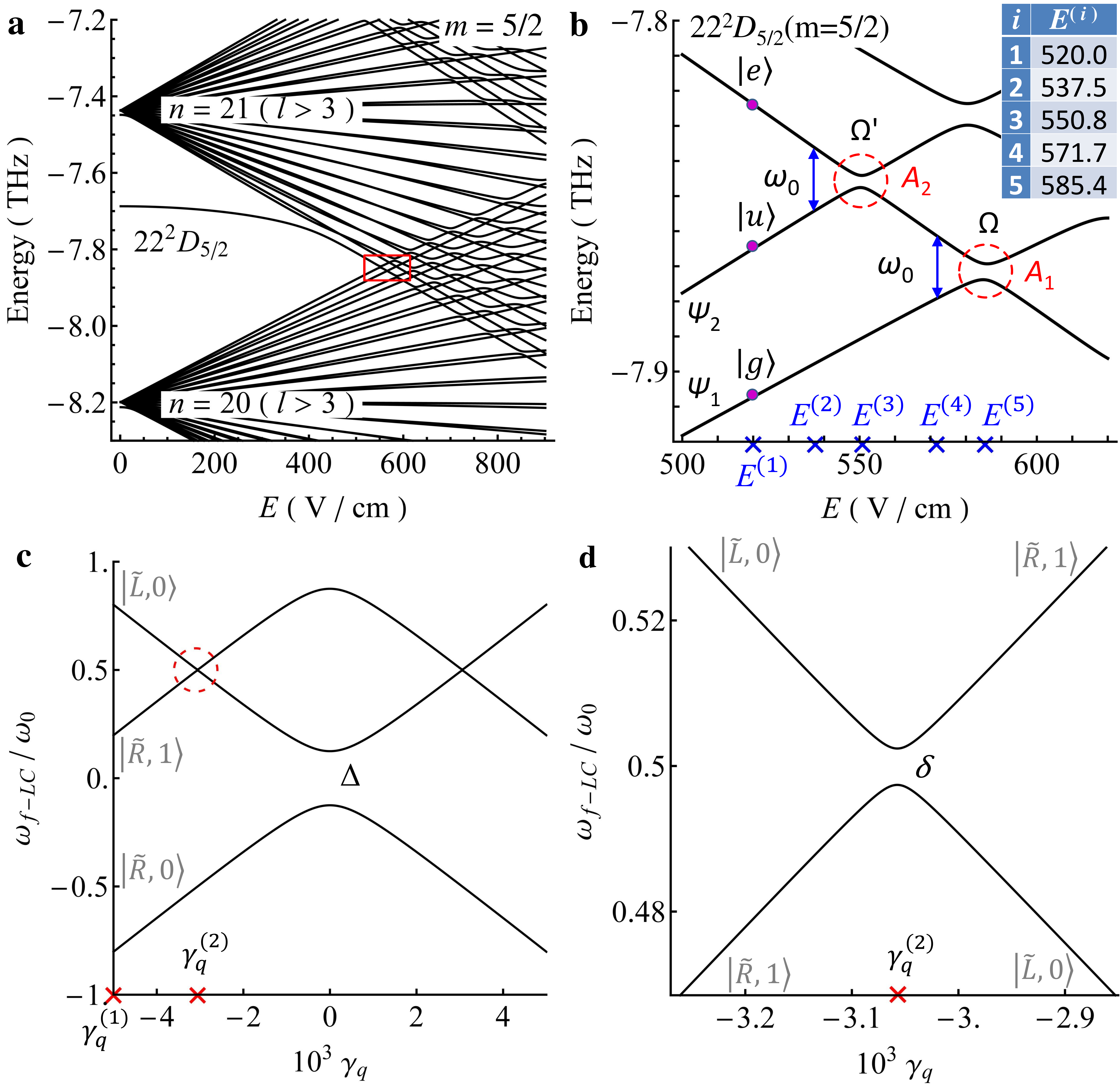}\\
\caption{(a) dc stark map of Rb around $22^{2}D_{5/2}$. The detail in rectangle frame is displayed in (b). The atom qubit is formed by the eigenstates $|e\rangle$ and $|g\rangle$ at $E^{(1)}=520.0$ V/cm on two adiabatic energy lines, which starts with $22^{2}D_{5/2}$ and the manifold state $|\psi_{1}\rangle$ composed of a set of $|n=20,l\geq3,j=l\pm\frac{1}{2},m=\frac{5}{2}\rangle$ at $E=0$. The eigenstate $|u\rangle$ on the curve starting with another manifold state $|\psi_{2}\rangle$, which is composed of a set of $|n=20,l\geq3,j=l\pm\frac{1}{2},m=\frac{5}{2}\rangle$, at $E=0$, is chosen to be an auxiliary state. Due to $|e\rangle-|u\rangle$ and $|e\rangle-|g\rangle$ interactions, two avoided crossings (labeled as $A_{1}$ and $A_{2}$ and energy separations of $\Omega$ and $\Omega'$) occur at $E^{(5)}$ and $E^{(3)}$. (c) Energy spectrum ($\omega_{f-LC}$) of the flux qubit-resonator interface vs. the parameter $\gamma_{q}$. At $\gamma^{(2)}_{q}=-5\times10^{-3}$, the flux qubit hardly interacts with the resonator. An anticrossing with a spacing of $\Delta=2\pi\times5$ GHz is located at $\gamma_{q}=0$. Another avoided crossing (spacing of $\delta=2\pi\times0.1$ GHz), induced by the resonant flux qubit-resonator coupling, happens at $\gamma^{(1)}_{q}=-3.06\times10^{-3}$. The detail is shown in (d).}\label{Fig2}
\end{figure}

Figure \ref{Fig2} (a) illustrates the energy spectrum of atom vs. the static electric $E$ field. It is seen that there exist plenty of avoided-level crossings. We focus on the ones, labeled as A1 and A2 in figure \ref{Fig2} (b), with the energy separations of $\Omega=2\pi\times4.6$ GHz and $\Omega'=2\pi\times3.2$ GHz. The anticrossing A2, located at $E^{(3)}=550.8$ V/cm, occurs between two adiabatic energy curves starting from $22^{2}D_{5/2}$ and a manifold $\psi_{2}$ state composed of a set of $|n=20,l\geq3,j=l\pm\frac{1}{2},m=\frac{5}{2}\rangle$ states at $E=0$. Here, $n$, $l$, $j$, and $m$ are the principle, orbital, total angular momentum, and magnetic quantum number of an atomic state, respectively. The anticrossing A1, happening at $E^{(5)}=585.4$ V/cm, is formed between the adiabatic energy curves starting from $\psi_{2}$ and another manifold $\psi_{1}$ state composed of a set of $|n=20,l\geq3,j=l\pm\frac{1}{2},m=\frac{5}{2}\rangle$ states at $E=0$.

The atomic qubit is formed by two $|e\rangle$ and $|g\rangle$ eigenstates at $E^{(1)}=520.0$ V/cm on the adiabatic curves associated with $22^{2}D_{5/2}$ and $\psi_{1}$ (see figure \ref{Fig2} (a)). The $|u\rangle$ eigenstate on the curve associated with $\psi_{2}$ is chosen as an auxiliary state. Adiabatic tuning $E$ varies the energy levels of $|\mu=e,g,u\rangle$, giving rise to the avoided crossings A1 and A2. Within the space consisting of $|e\rangle$, $|g\rangle$, and $|u\rangle$, the Hamiltonian of atom is given by
\begin{equation}
H_{a}=\sum_{\mu=e,g,u}\hbar\omega_{\mu}|\mu\rangle\langle\mu|+\frac{\hbar\Omega}{2}(|e\rangle\langle g|+|g\rangle\langle e|)+\frac{\hbar\Omega'}{2}(|e\rangle\langle u|+|u\rangle\langle e|),
\end{equation}
where the electric-field-dependent energies $\omega_{\mu=e,g,u}$ of atomic states are derived as $\omega_{e}=-7.81-(E-500)\times7.3\times10^{-4}$ THz, $\omega_{g}=-7.92+(E-500)\times5.5\times10^{-4}$ THz, and $\omega_{u}=-7.88+(E-500)\times6.3\times10^{-4}$ THz. The electric $E$ field is in the unit of V/cm and within the range from 500 V/cm to 620 V/cm. The atom-resonator interaction is expressed as
\begin{equation}
V_{a}=\frac{\hbar g_{a}}{2}(b^{\dag}|g\rangle\langle e|+|e\rangle\langle g|b)+\frac{\hbar g'_{a}}{2}(b^{\dag}|u\rangle\langle e|+|e\rangle\langle u|b),
\end{equation}
with the coupling strengths $g_{a}=|\langle e|D|g\rangle|{\cal{E}}/\hbar=2\pi\times1.0$ GHz and $g'_{a}=|\langle e|D|u\rangle|{\cal{E}}/\hbar=2\pi\times0.5$ GHz. The atom resonantly interacts with the resonator at $E^{(2)}=537.5$ V/cm ($\omega_{e}-\omega_{u}=\omega_{0}$) and $E^{(4)}=571.7$ V/cm ($\omega_{e}-\omega_{g}=\omega_{0}$).

Besides interacting with the atom, the $LC$ resonator is also inductively coupled to a three-JJ flux qubit (see figure \ref{Fig1} (a)). The mutual inductance $M$ between the flux qubit and the inductor of resonator depends on their geometries and the relative positioning and can be as high as tens of pH~\cite{Science:Chiorescu2003,PRB:Wang2010,PRL:Johnson2012,NewJPhys:Yamamoto2014}. We choose $M=27$ pH to obtain a strong enough flux qubit-resonator coupling. The flux qubit is biased by an external magnetic flux $\Phi_{ex}$ to tune the frequency spacing $\varepsilon=2I_{p}\Phi_{0}\gamma_{q}/\hbar$ (the qubit maximum persistent current $I_{p}=0.8$ $\mu$A~\cite{Science:Chiorescu2003}, the flux quantum $\Phi_{0}=h/(2e)$, and the phase-bias parameter $\gamma_{q}=\Phi_{ex}/\Phi_{0}-1/2$) between the ground $|R\rangle$ (clockwise) and upper $|L\rangle$ (anticlockwise) states. The flux-qubit Hamiltonian is written as
\begin{equation}
H_{f}=-\frac{\hbar\varepsilon}{2}\sigma_{f,z}-\frac{\hbar\Delta}{2}\sigma_{f,x},
\end{equation}
with the Pauli matrices $\sigma_{f,z}=|\tilde{L}\rangle\langle\tilde{L}|-|\tilde{R}\rangle\langle\tilde{R}|$, $\sigma_{f,x}=\sigma^{\dag}_{f,-}+\sigma_{f,-}$, $\sigma_{f,-}=|\tilde{R}\rangle\langle\tilde{L}|$, and the tunnel splitting $\Delta=2\pi\times5.0$ GHz. The flux qubit-resonator interaction is given by
\begin{equation}
V_{f}=-\hbar g_{f}(b^{\dag}+b)\sigma_{f,z},
\end{equation}
with the coupling strength $g_{f}=(MI_{p}/\hbar)\sqrt{\hbar\omega_{0}/(2L)}=2\pi\times0.2$ GHz. Figure \ref{Fig2} (c) displays the energy spectrum of the flux qubit-resonator interface, i.e., \begin{equation}
(H_{LC}+H_{f}+V_{f})|\Psi_{f-LC}\rangle=\hbar\omega_{f-LC}|\Psi_{f-LC}\rangle,
\end{equation}
where $\omega_{f-LC}$ is the eigenvalue, the eigenstate $|\Psi_{f-LC}\rangle$ is spanned in the basis of $\{|o,n_{p}\rangle,o=\tilde{L},\tilde{R};n_{p}=0,1,2,...\}$, and $n_{p}$ denotes the microwave photon number, vs. the parameter $\gamma_{q}$. At $\gamma^{(1)}_{q}=-5\times10^{-3}$, the resonator is hardly coupled to the flux qubit. The strong interaction between $|\tilde{R}\rangle$ and $|\tilde{L}\rangle$ gives rise to an avoided crossing with a spacing of $\Delta$ at $\gamma_{q}=0$. The resonant flux qubit-resonator coupling happens at $\gamma^{(2)}_{q}=-3.06\times10^{-3}$, where an anticrossing occurs between $|\tilde{R},1\rangle$ and $|\tilde{L},0\rangle$ with an energy separation of $\delta=\frac{2V_{f}\Delta}{\hbar\sqrt{\varepsilon^{2}+\Delta^{2}}}=2\pi\times0.1$ GHz (see figure \ref{Fig2} (d)).

For the whole hybrid system, the Hilbert space is spanned by $\{|o,n_{p},\mu\rangle,o=\tilde{L},\tilde{R};n_{p}=0,1;\mu=e,g,u,s\}$. The extra atomic $|s\rangle$ state is introduced to simulate the decay of atomic states $|\mu=e,g,u\rangle$. The system dynamics is governed by the master equation
\begin{equation}\label{MasterEq}
\dot{\rho}=-i[H/\hbar,\rho]+\gamma_{relax}{\cal{D}}[\sigma_{f,-}]\rho+\frac{\gamma_{\phi}}{2}{\cal{D}}[\sigma_{f,z}]\rho+\kappa{\cal{D}}[b]\rho+\Gamma\sum_{\mu=e,g,u}{\cal{D}}[|s\rangle\langle \mu|]\rho,
\end{equation}
where the density matrix operator $\rho$, the system Hamiltonian
\begin{equation}
H=H_{LC}+H_{a}+H_{f}+V_{a}+V_{f},
\end{equation}
and the Lindblad superoperator
\begin{equation}
{\cal{D}}[K]\rho=K\rho K^{\dag}-\frac{1}{2}K^{\dag}K\rho-\frac{1}{2}\rho K^{\dag}K.
\end{equation}
$\gamma_{relax}=2\pi\times0.03$ MHz and $\gamma_{\phi}=2\pi\times0.1$ MHz are the energy relaxation and pure dephasing rates of the flux qubit~\cite{PRL:Stern2014}. $\kappa=\omega_{0}/Q=2\pi\times0.2$ MHz (the factor $Q=10^{5}$) is the loss rate of intraresonator energy. Since the atom in Rydberg state is close to the capacitor surface, the stray field from the surface strongly reduces the lifetimes of Rydberg states~\cite{PRA:Tauschinsky2010,PRA:HermannAvigliano2014,PRL:Chan2014}. Here, for simplicity, we have assumed that all Rydberg states $|\mu=e,g,u\rangle$ have the same decay rate $\Gamma$, which is estimated to be $2\pi\times0.15$ MHz~\cite{PRA:Crosse2010}. The Casimir-Polder shifts of Rydberg states due to the stray field are small and have been included in $\omega_{\mu=e,g,u}$. For convenience, we list main physical parameters and their corresponding values in Table 1.

\subsection{Preparation of entangled quantum states}

Preparing different entangled states is important in quantum information processing~\cite{PRA:Cabrillo1999,PRA:Kraus2008,PRL:Lin2016}. Based on our hybrid system, this can be accomplished by varying the electrostatic field $E$ and the magnetic flux bias $\Phi_{ex}$, i.e., the parameter $\gamma_{q}$. Here we focus on the maximally entangled three-qubit states~\cite{PRA:Dur2000}, the so-called Greenberger-Horne-Zeilinger (GHZ) state,
\begin{equation}
|GHZ\rangle=\frac{1}{\sqrt{2}}(|\tilde{L},1,e\rangle+|\tilde{R},0,g\rangle),
\end{equation}
and the W state,
\begin{equation}
|W\rangle=\frac{1}{\sqrt{3}}(|\tilde{L},0,g\rangle+|\tilde{R},1,g\rangle+|\tilde{R},0,e\rangle).
\end{equation}
The W state retains the bipartite entanglement when one qubit is traced out.

\begin{figure}
\centering
\includegraphics[width=10.0cm]{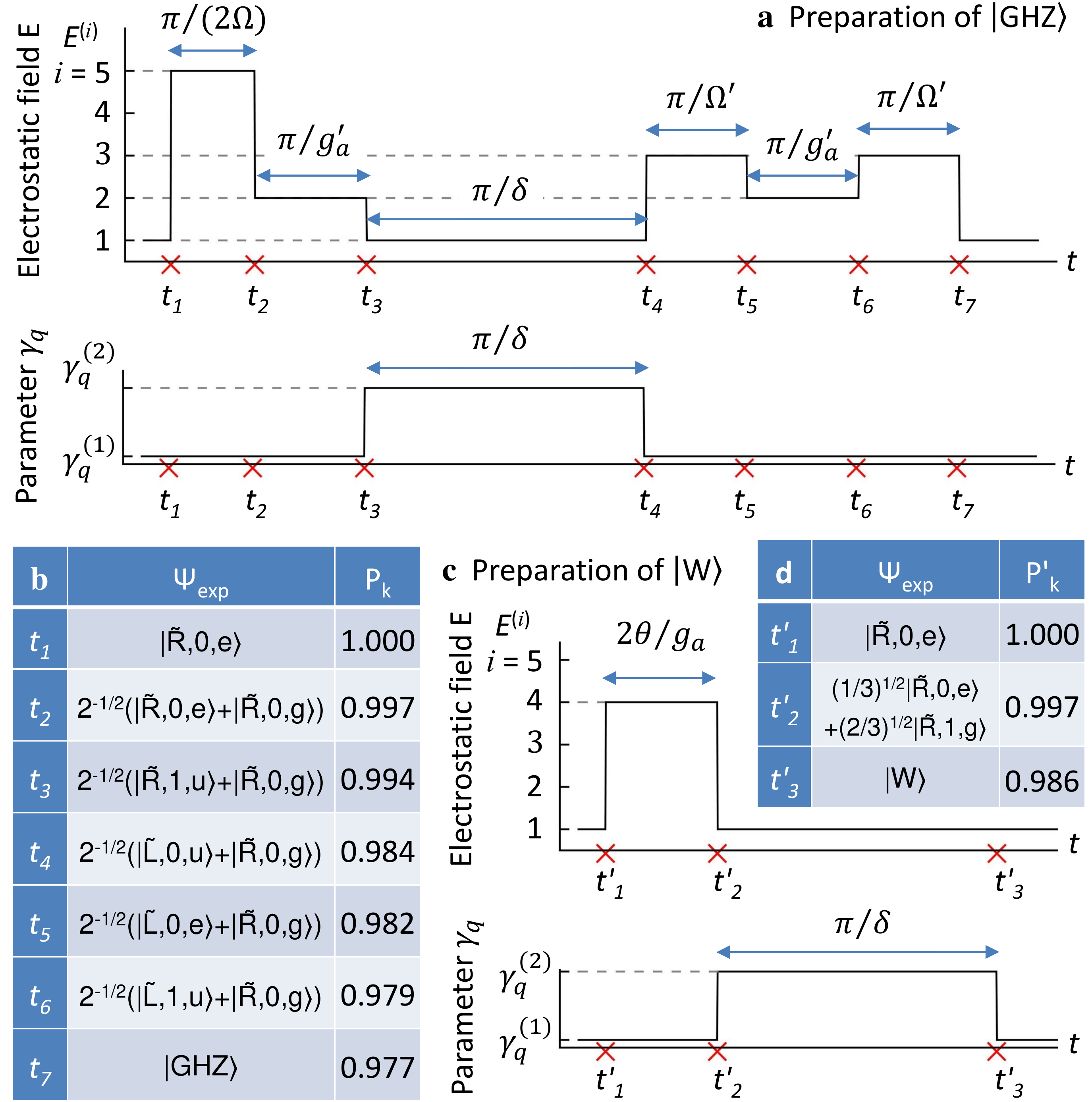}\\
\caption{(a) Preparation of $|GHZ\rangle$. The electric field $E$ and parameter $\gamma_{q}$ are sequentially swept among different gears, $\{E^{(i)},i=1,...,5\}$ and $\{\gamma^{(i)}_{q},i=1,2\}$. At different times $\{t_{i},i=1,...,7\}$, the hybrid system is expected to stay at certain states $|\Psi_{exp}\rangle$ listed in (b). The probability of the system in $|\Psi_{exp}\rangle$, $\{P_{i}=\langle\Psi_{exp}|\rho(t_{i})|\Psi_{exp}\rangle,i=1,...,7\}$, are obtained by solving the master equation~(\ref{MasterEq}). (c) Sweeping processes of $E$ and $\gamma_{q}$ for preparing $|W\rangle$. The corresponding expected states $|\Psi_{exp}\rangle$ and probabilities $\{P'_{i}=\langle\Psi_{exp}|\rho(t'_{i})|\Psi_{exp}\rangle,i=1,2\}$ at certain times $\{t'_{i},i=1,2\}$ are listed in (d).}\label{Fig3}
\end{figure}

The preparation of $|GHZ\rangle$ is implemented via the following five steps (see figure \ref{Fig3} (a)): (1) Initially, $E$ and $\gamma_{q}$ are set at $E^{(1)}$ and $\gamma^{(1)}_{q}$, where both atom and flux qubit are far-off-resonantly coupled to the resonator, and the hybrid system is in the pure $|R,0,e\rangle$ state. Then, $E$ is rapidly increased to $E^{(5)}$, where the anticrossing $A_{1}$ is situated (see figure \ref{Fig2} (b)), and stays at $E^{(5)}$ for a time duration of $\pi/(2\Omega)$. As a result, the atom transits to the superposition $(|e\rangle+|g\rangle)/\sqrt{2}$ state. (2) $E$ is reduced to $E^{(2)}$ nonadiabatically, switching on the resonant interaction between the microwave resonator and the atomic $|e\rangle-|u\rangle$ transition. After a $\pi$-pulse time length of $\pi/g'_{a}$, $E$ is further decreased back to $E^{(1)}$ fast. In this step, the $|e\rangle$ component of atomic state evolves to the auxiliary $|u\rangle$ state with a microwave photon being injected into the resonator. (3) $\gamma_{q}$ is changed to $\gamma^{(2)}_{q}$ rapidly, turning on the resonant flux qubit-resonator interaction. This resonant coupling lasts for a time duration of $\pi/\delta$, completing the energy quantum transfer from the resonator to the flux qubit, and the flux qubit flips its state. (4) $E$ goes up to $E^{(3)}$ quickly and stays for a time length of $\pi/\Omega'$. The $|u\rangle$ component of atomic state evolves back to $|e\rangle$. (5) Repeat steps (2) and (4) sequentially. Finally, the hybrid system arrives at $|GHZ\rangle$.

Solving the master equation~(\ref{MasterEq}) gives the probabilities of the hybrid system in different states. The GHZ state is produced within 8 ns, much shorter than any decoherence times, with a state-preparation fidelity of 0.977 (see figure \ref{Fig3} (b)). Besides the inevitable qubit decays, the fidelity is limited by the property of dc Stark map of atom. For example, the avoided crossings $A_{1}$ and $A_{2}$ are not separated far away from each other, leading to the mixture of $|\mu=e,g,u\rangle$. In addition, the locations of resonant atom-resonator coupling, $E^{(2)}$ and $E^{(4)}$, are not very far away from that of corresponding anticrossings, $E^{(3)}$ and $E^{(5)}$, resulting in the incomplete energy transfer between atom and resonator.

The preparation of $|W\rangle$ is straightforward (see figure \ref{Fig3} (c)). Again, we start with the initial $|\tilde{R},0,e\rangle$ state and conditions of $E=E^{(1)}$ and $\gamma_{q}=\gamma^{(1)}_{q}$. Then, $E$ is increased to $E^{(4)}$ nonadiabatically, where the resonator is resonantly coupled to the atomic $|e\rangle-|g\rangle$ transition. After staying at $E^{(4)}$ for a time length of $2\theta/g_{a}$ with $\theta=\textrm{sin}^{-1}\sqrt{2/3}$, $E$ returns to $E^{(1)}$ fast. The system state evolves to $\sqrt{2/3}|\tilde{R},1,g\rangle+\sqrt{1/3}|\tilde{R},0,e\rangle$. Next, $\gamma_{q}$ goes up to $\gamma^{(2)}_{q}$ rapidly, stays for a $\pi$-pulse duration of $\pi/\delta$, and then is reduced back to $\gamma^{(1)}_{q}$ quickly. Finally, we obtain the entangled $|W\rangle$ state. The state preparation is completed within 6 ns with a fidelity of 0.986 (see figure \ref{Fig3} (d)).

\subsection{Three-qubit logic operation}

Besides the preparation of entangled states, three-qubit logic operations can be also performed on this hybrid system. Here, as an example, we focus on the Toffoli gate, which is the key component for the quantum-error correction~\cite{Book:Mermin2007} and so far has been implemented only on trapped ions~\cite{PRL:Monz2009} and superconducting circuits~\cite{Nature:Fedorov2012}. In our performance, the circulating persistent-current and microwave-photon states play the joint control role while the atom acts as the target qubit. The atom flipping between $|e\rangle$ and $|g\rangle$ is implemented via periodically alternating the electric field $E$ around a central value $E^{(l)}$ with an amplitude of $\delta E$ and at a rate of $\omega_{l}$. This extra atom-field interaction is described by
\begin{equation}
V_{l}=\hbar g_{l}(|e\rangle\langle g|+|g\rangle\langle e|)\cos\omega_{l}t,
\end{equation}
with the coupling strength $g_{l}=|\langle e|D|g\rangle|\delta E/\hbar$. The realization of Toffoli gate requires that the atomic-qubit $|e\rangle-|g\rangle$ transition with the flux qubit-resonator combined state $|\tilde{L},1\rangle$ states can be distinguished from that with other combined states, i.e., $|\tilde{R},0\rangle$, $|\tilde{R},1\rangle$, and $|\tilde{L},0\rangle$, in spectrum. The gate operation is performed simply via setting $\omega_{l}$ resonantly to the $|\tilde{L},1,e\rangle-|\tilde{L},1,g\rangle$ transition.

\begin{figure}
\centering
\includegraphics[width=10.0cm]{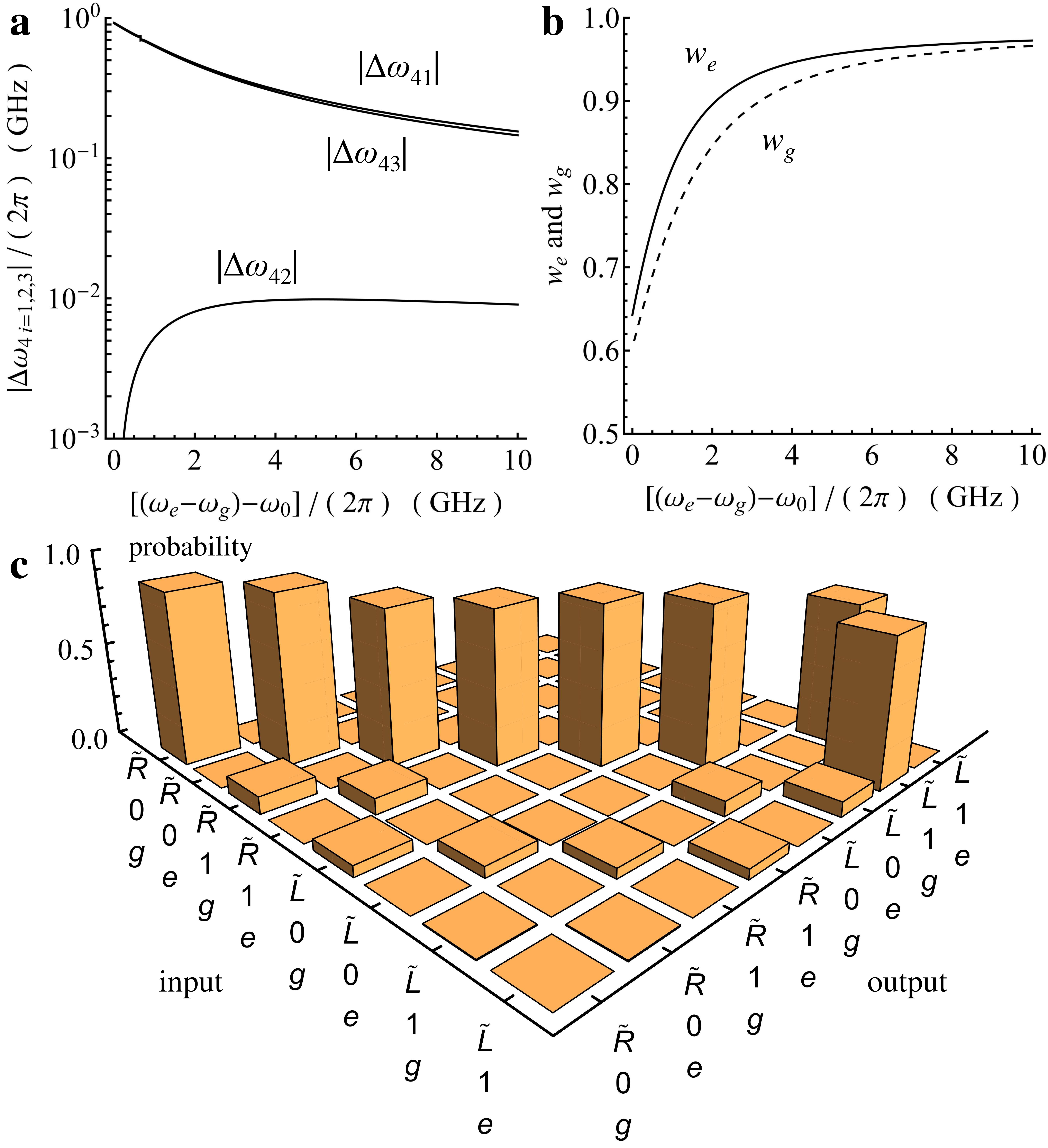}\\
\caption{(a) Frequency differences $|\Delta\omega_{4i}|=|\omega_{4}-\omega_{i}|$ ($i=1,2,3$) vs. $[(\omega_{e}-\omega_{g})-\omega_{0}]$ with $\varepsilon=2\pi\times21.0$ GHz (corresponding to $\gamma_{q}=\gamma^{(l)}_{q}=-4.3\times10^{-3}$). (b) Weights $w_{e}$ and $w_{g}$ of $|\tilde{L},1,e\rangle$ and $|\tilde{L},1,g\rangle$ components in $|\tilde{L},1,e\rangle'$ and $|\tilde{L},1,g\rangle'$, respectively, as a function of $[(\omega_{e}-\omega_{g})-\omega_{0}]$. (c) Truth table amplitudes of the Toffoli gate. The average probability reaches 82.6\%.}\label{Fig4}
\end{figure}

For a pure state, such as $|\tilde{R},0,g\rangle$, due to the off-resonant flux qubit-resonator and atom-resonator interactions, $|\tilde{R},0,g\rangle$ is slightly mixed with other pure states, resulting in an energy-level shifted $|\tilde{R},0,g\rangle'$ state in which the $|\tilde{R},0,g\rangle$ component weights predominantly. We use $\omega_{1}$, $\omega_{2}$, $\omega_{3}$, and $\omega_{4}$ to respectively denote the frequencies of shifted $|\tilde{R},0,e\rangle'-|\tilde{R},0,g\rangle'$, $|\tilde{R},1,e\rangle'-|\tilde{R},1,g\rangle'$, $|\tilde{L},0,e\rangle'-|\tilde{L},0,g\rangle'$, and $|\tilde{L},1,e\rangle'-|\tilde{L},1,g\rangle'$ transitions. When $|(\omega_{e}-\omega_{g})-\omega_{0}|\gg g_{a}$ and $|\varepsilon-\omega_{0}|\gg g_{f}$, these four transitions become degenerate. Figure \ref{Fig4} (a) displays the frequency differences $\Delta\omega_{4i}=\omega_{4}-\omega_{i}$ ($i=1,2,3$) as a function of $(\omega_{e}-\omega_{g})$ with a fixed $\varepsilon$. As one can see, the off-resonant intersubsystem couplings lead to the discrimination of four transitions. At $(\omega_{e}-\omega_{g})=\omega_{0}+2\pi\times7$ GHz, we have $|\Delta\omega_{42}|\approx2\pi\times10$ MHz and $|\Delta\omega_{41}|\approx|\Delta\omega_{43}|\approx2\pi\times0.2$ GHz and the weights $w_{e}=|\langle\tilde{L},1,e|\tilde{L},1,e\rangle'|^{2}\approx0.96$ and $w_{g}=|\langle\tilde{L},1,g|\tilde{L},1,g\rangle'|^{2}\approx0.95$ (see figure \ref{Fig4} (b)). Setting $\omega_{l}$ equal to $\omega_{4}$ and choosing $g_{l}<|\Delta\omega_{4i}|$ ($i=1,2,3$), the oscillating electric $E$ field resonantly drives the $|\tilde{L},1,e\rangle'-|\tilde{L},1,g\rangle'$ transition but hardly affects other three transitions due to the off-resonant interaction.

The Toffoli gate is performed as following: Initially, $E$ and $\gamma_{q}$ are set at $E^{(1)}$ and $\gamma^{(1)}_{q}$, i.e., $(\omega_{e}-\omega_{g})$ and $\varepsilon$ are far away from $\omega_{0}$. Then, $E$ and $\gamma_{q}$ are rapidly tuned to $E^{(l)}=564.4$ V/cm and $\gamma^{(l)}_{q}=-4.3\times10^{-3}$, where $(\omega_{e}-\omega_{g})=\omega_{0}+2\pi\times7.0$ GHz and $\varepsilon=\omega_{0}+2\pi\times21.0$ GHz. Next, $E$ starts oscillating around $E^{(l)}$ with an amplitude of $\delta E=0.7$ V/m satisfying $g_{l}=2\pi\times7.0$ MHz. After a $\pi$-pulse time duration of $\pi/g_{l}$, $E$ and $\gamma_{q}$ are changed back to $E^{(1)}$ and $\gamma^{(1)}_{q}$ nonadiabatically. Figure \ref{Fig4} (c) shows the resulting register populations of the Toffoli operation derived by solving the master equation. It is seen that the quantum logic gate preserves the atom-qubit state when the flux qubit and resonator are in $|\tilde{R},0\rangle$, $|\tilde{R},1\rangle$, and $|\tilde{L},0\rangle$, whereas the atom-qubit state switches between $|e\rangle$ and $|g\rangle$ with a high probability for $|\tilde{L},1\rangle$. On average the probability reaches 82.6\%. Here we should note that increasing the flux qubit-resonator coupling strength $g_{f}$ can efficiently increase the frequency differences $|\Delta\omega_{4i}|$ ($i=1,2,3$) with slightly reducing the weights $w_{e}$ and $w_{g}$. As a result, the fidelity of gate operation is enhanced. However, this relies on a larger mutual inductance $M$, challenging the experimental feasibility.

\section{Conclusion}

We have investigated a hybrid system composed of an $LC$ resonator inductively interacting with a flux qubit and being capacitively coupled to a Rydberg atom. The qubit flipping and intercomponent interactions are controlled by the external magnetic flux $\Phi_{ex}$ and electrostatic field $E$. Fully entangled states, $|GHZ\rangle$ and $|W\rangle$, are prepared via simply tuning $\Phi_{ex}$ and $E$, and the three-qubit logic gates can be also implemented. Our hybrid system brings three different two-level systems together and provides a platform for transferring information among superconducting, photonic, and atomic realizations of quantum computing. The recent experimental progresses in the flux qubit-resonator coupling \cite{Nature:Chiorescu2004} and Rydberg atoms interacting with a microwave cavity~\cite{PRL:Hogan2012,arxiv:Stammeier} indicate the prospect of three-qubit entanglement and logic gate operation discussed in this work.

\section*{Acknowledgements}
This research has been supported by the National Research Foundation Singapore \& by the Ministry of Education Singapore Academic Research Fund Tier 2 (Grant No. MOE2015-T2-1-101).

\begin{table}
\fontsize{10}{15}\selectfont
\centering
\begin{tabular}{l l l l l}
\hline
Physical parameters (units)~~~~~~~~~~~~~~~~~~~~~~&Symbols~~~~~~~& Values \\
\hline
Capacitance (aF) & $C$ & 256 \\
Inductance (nH) & $L$ & 247 \\
Resonator frequency (GHz) & $\omega_{0}/(2\pi)$ & 20 \\
Static electric field (V/cm) & $E$ & \\
Specific values of electric field (V/cm) & $E^{(i=1,...,5)}$ & \\
Anticrossing A1 (GHz) & $\Omega/(2\pi)$ & 4.6 \\
Anticrossing A2 (GHz) & $\Omega'/(2\pi)$ & 3.2 \\
Energies of atomic states $|\mu=e,g,u\rangle$ & $\omega_{\mu=e,g,u}$ & \\
Atom ($|g\rangle-|e\rangle$)-resonator coupling (GHz) & $g_{a}/(2\pi)$ & $1.0$ \\
Atom ($|u\rangle-|e\rangle$)-resonator coupling (GHz) & $g'_{a}/(2\pi)$ & $0.5$ \\
Flux-bias parameter & $\gamma_{q}$ & \\
Specific values of flux-bias parameter & $\gamma^{(i=1,2)}_{q}$ & \\
Flux-qubit frequency spacing (GHz) & $\epsilon$ & \\
Tunnel splitting of flux qubit (GHz) & $\Delta/(2\pi)$ & $5.0$ \\
Mutual inductance (pH) & $M$ & 27 \\
Flux qubit-resonator coupling (GHz) & $g_{f}/(2\pi)$ & $0.2$ \\
Anticrossing between $|R,1\rangle$ and $|L,0\rangle$ (GHz) & $\delta/(2\pi)$ & 0.1 \\
Relaxation rate of flux qubit (MHz) & $\gamma_{relax}/(2\pi)$ & 0.03 \\
Pure dephasing rate of flux qubit (MHz) & $\gamma_{\phi}/(2\pi)$ & 0.1 \\
Loss rate of resonator (MHz) & $\kappa/(2\pi)$ & 0.2 \\
Decay rate of Rydberg states (MHz) & $\Gamma/(2\pi)$ & 0.15 \\
Central value of oscillating field (V/cm) & $E^{(l)}$ & 564.4 \\
Amplitude of oscillating field (V/m) & $\delta E$ & 0.7 \\
Frequency of oscillating field (GHz) & $\omega_{l}/(2\pi)$ & \\
Atom-field coupling in Toffoli gate (MHz) & $g_{l}/(2\pi)$ & $7.0$ \\
\hline
& & \\
\end{tabular}
\caption{List of main physical parameters in the three-qubit hybrid structure.}\label{Table}
\end{table}

\end{document}